%% file: paper.tex
\documentclass[11pt]{article}
\usepackage{graphicx}
\usepackage{epsfig}
 
\long\def\inst#1{\par\nobreak\kern 4pt\nobreak
    {\it #1}\par\vskip 10pt plus 3pt minus 3pt}

\large
 
\def\Thebibliography#1{\section*{REFERENCES}\list
{[\arabic{enumi}]}{\settowidth\labelwidth{[#1]}\leftmargin\labelwidth
 \advance\leftmargin\labelsep
 \usecounter{enumi}}
 \def\newblock{\hskip .11em plus .33em minus .07em}
 \sloppy\clubpenalty4000\widowpenalty4000
 \sfcode`\.=1000\relax}

\def\Journal#1&#2&#3(#4){\unskip, #1~{\bf #2} (#4) #3}

\def\PLB{Phys.\ Lett.\ {\bf B}}

\def\PRD{Phys.\ Rev.\ {\bf D}}

\def\ZPC{Z.\ Phys.\ {\bf C}}

\def\babar{\mbox{\sl B\hspace{-0.4em} {\small\sl A}\hspace{-0.37em} \sl B\hspace{-0.4em} {\small\sl A\hspace{-0.02em}R}}}
 
\def\pep2{PEP-II}

\def\fbi{fb$^{-1}$\ }

\def\dstophipi{ ${\rm D_s^{\pm}}\rightarrow \phi\pi^{\pm}$ }

\def\Dphipi{${\rm D^{\pm}} \rightarrow {\rm \phi \pi^{\pm}}$}

\def\etal{{\it et al.}}

\def\epem{$e^{+}e^{-}$ }

\def\ups{$\Upsilon$(4S) }
\def\bbbar{${\rm B\overline{B}}$ }
\def\b0b0{${\rm B^{0}\overline{B^0}}$}

\def\fbi{fb$^{-1}$}

\def\BDsXc{${\rm B} \rightarrow {\rm D^{*+}_s X_c}$}
\def\BDstarDsstar{${\rm B^{0}} \rightarrow {\rm D^{*-} D_s^{*+}}$}
\def\BDstarDs{${\rm B^{0}} \rightarrow {\rm D^{*-} D_s^{+}}$}
\def\BDstarDss{${\rm B^{0}} \rightarrow {\rm D^{*-} D_s^{(*)+}}$}

\def\btou{{\rm b\rightarrow u}}
\def\btoccs{{\rm b\rightarrow c\overline{c}s}}

\def\VuboVcb{{\rm V_{ub}/V_{cb}}}

\def\BN{{\rm B^0}}

\def\Dstarpm{{\rm D^{*\pm}}}

\def\DN{{\rm D^0}}

\def\Dpm{{\rm D^\pm}}

\def\Dspm{{\rm D_s^\pm}}
\def\Dspmstar{{\rm D_s^{*\pm}}}
\def\Dspms{{\rm D_s^{(*)\pm}}}
\def\Dsphipi{{\rm D_s^{\pm}} \rightarrow {\rm \phi \pi^{\pm}} }

\def\ev   {${\rm \,e\kern -0.08em V}$}
\def\kev  {${\rm \,ke\kern -0.08em V}$} 
\def\mev  {${\rm \,Me\kern -0.08em V}$} 
\def\gev  {${\rm \,Ge\kern -0.08em V}$} 
\def\gevc {${{\rm \,Ge\kern -0.08em V\!/}c}$} 
\def\mevc {${{\rm \,Me\kern -0.08em V\!/}c}$} 
\def\gevcc{${{\rm \,Ge\kern -0.08em V\!/}c^2}$} 
\def\mevcc{${{\rm \,Me\kern -0.08em V\!/}c^2}$} 
\def\mum  {${\,\mu\rm m}$} 
\newcommand{\dedx}{${\mathrm{d}\hspace{-0.1em}E/\mathrm{d}x}$}

\newcommand{\be}{\begin{equation}}
\newcommand{\ee}{\end{equation}}
\newcommand{\ba}{\begin{array}{c}}
\newcommand{\ea}{\end{array}}
\newcommand{\beqn}{\begin{eqnarray}}
\newcommand{\eeqn}{\end{eqnarray}}

\newcommand{\twd}{\textwidth}

\newcommand{\BABARPubYear}    {00}

\newcommand{\BABARConfNumber} {13}
\newcommand{\SLACPubNumber} {8535}

\begin{document}
{\pagestyle{empty}

\begin{flushright}
\babar-CONF-\BABARPubYear/\BABARConfNumber \\
SLAC-PUB-\SLACPubNumber
\end{flushright}

\par\vskip 3cm

\begin{center}
\Large \bf \boldmath
Study of inclusive $\Dspms $ production in $B$ decays and measurement of
\BDstarDss \ decays using a partial reconstruction technique
\end{center}
\bigskip

\begin{center}
\large The \babar\ Collaboration\\
\mbox{ }\\
July 25, 2000
\end{center}
\bigskip \bigskip

\begin{center}
\large \bf Abstract
\end{center}
Electron-positron annihilation data collected by the \babar\
detector near the
\ups resonance are used to study the inclusive decay of B mesons to 
$\Dspm $ and $\Dspmstar$ mesons, where the $\Dspm $ is reconstructed using 
the decay $\Dsphipi$. 
The production fraction of inclusive $\Dspms $ and the corresponding momentum
spectra have been determined. The exclusive decays \BDstarDss \ are 
observed with a partial reconstruction technique which uses the soft
pion from the $\Dstarpm $ decay in association with the reconstructed 
$\Dspms $. The beam
energy constraint is used to determine the missing mass recoiling against 
the $\Dspm $ system, showing a clear signal for this process. 
From the observed 
rates, preliminary results for the corresponding
branching fractions have been obtained.

\vfill
\begin{center}
Submitted to the XXX$^{th}$ International 
Conference on High Energy Physics, Osaka, Japan.
\end{center}

\newpage
}

\input pubboard/authors

\setcounter{footnote}{0}

\section{Introduction}

\begin{figure}[htbp]
\begin{picture}(300,100)
\put(55,90){\includegraphics{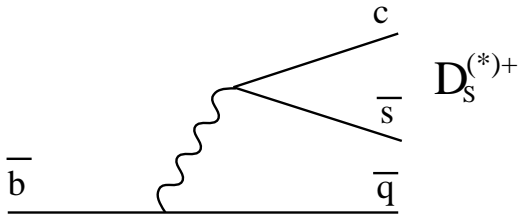}}
\put(340,-10){\makebox(60,40)[b]{\begin{minipage}[b]{50mm}
\protect\caption{{\small
The main spectator diagram leading to the production of $\Dspms$ 
mesons in B decays.}
\protect\label{fig:dsprod}}
\end{minipage}}}
\end{picture}
\end{figure}

The production of the $\Dspms$ meson in B decays allows one to study the mechanisms
leading to the creation of a $\rm{ c \bar{s}}$ quark pair.
The main diagram contributing to this decay is shown
in Fig.~\ref{fig:dsprod}. 
Other B decay diagrams also contribute, although at a lower level,
but no attempt is made to quantify their rate in this paper.
As has been pointed out in Ref.~\cite{FWD:1},
the $\btoccs$ decay rate 
may be large and could help to explain the theoretical
difficulties \cite{bigi:1} 
in accounting simultaneously for the total inclusive B decay rate and the 
semileptonic branching fraction of the B meson. 
As a longer term goal, the measurement of the rate and momentum spectrum 
of $\Dspm$ meson
produced in B decays beyond the kinematic limit for the process \BDsXc \ 
could be used to study $\btou$ transitions. 
Despite the fact that purely hadronic final states are more difficult to
understand theoretically, one may use the particular decay described in this
paper to extract $\VuboVcb$ \cite{AZLOPR:1}.
This document reports 
measurements made with the \babar \ detector of both the inclusive $\Dspms$
production rates and momentum spectra in B decays and the branching fractions 
of two specific two-body B decay modes involving a $\Dspms$ meson. 
The latter measurements are made using a partial 
reconstruction technique.

\section{The detector and the data sample}

A description of the \babar \ detector and the definition
of many general analysis procedures can be found in an accompanying 
paper \cite{babdet}.  Here only the components of the detector
most crucial to this analysis are briefly summarized.

Charged particles are detected and their momenta measured by a
combination of a central drift chamber (DCH) with a helium-based
gas and a five-layer (double-sided) silicon vertex
tracker (SVT), embedded in a 1.5 T solenoidal field produced by a 
superconducting magnet.  The charged particle momentum resolution
is approximately $(\delta p_T/p_T)^2 = (0.0015 \, p_T)^2 + (0.005)^2$,
where $p_T$ is in GeV$/c$.  The SVT, with typically 10\mum\ single-hit
resolution, provides vertex information in both the transverse plane 
and in $z$.

Particles are identified using a combination of measurements from
all the \babar \ components.  Charged particle identification
exploits ionization energy loss measured in the DCH and SVT
as well as Cherenkov radiation measured in a ring imaging
detector (DIRC).  Electrons and photons are identified by the CsI 
electromagnetic calorimeter.


Multihadronic events produced in \epem annihilation at the \pep2 collider 
(SLAC) and collected with the \babar \ detector have been used in this 
analysis.  These data were taken at the  \ups resonance center of mass 
energy  and at an energy about 40\mev\ below the \bbbar threshold. 
The integrated luminosity for on resonance data is 7.73 \fbi \ and
1.17 \fbi \ for off resonance.

\section{Inclusive {\boldmath $\Dspm$} production}

\subsection{{\boldmath $\Dspm$} reconstruction}

The $\Dspm$ mesons are reconstructed using the mode \dstophipi
with $\phi \rightarrow {\rm K^+ K^-}$.
In order to obtain a
sufficiently clean sample, particle identification is necessary. 
To this end, both energy loss (\dedx) information from
the Drift Chamber and the Vertex Detector and the DIRC
(a Cherenkov imaging detector) are used 
to identify the kaons produced in the $\phi$ decay.

The selection is based on the likelihoods given 
by each detector and uses, for each track, the ratio of 
likelihoods for the pion and the kaon mass hypotheses $L_\pi/L_K$.
If this ratio is less than unity for at least one of the considered 
subsystems, the particle is selected as a kaon. 
The DIRC is used both in the positive identification mode and the veto mode. 
A tighter level of identification is also available using
a total likelihood defined
as the product of the likelihoods of each subsystem.
In this case the track is tagged as a kaon if the ratio of the total
likelihoods for the pion and kaon mass hypotheses is less than unity.

\begin{center}
\begin{figure} [t]
\begin{minipage}{.45\twd}
\includegraphics[width=\twd]{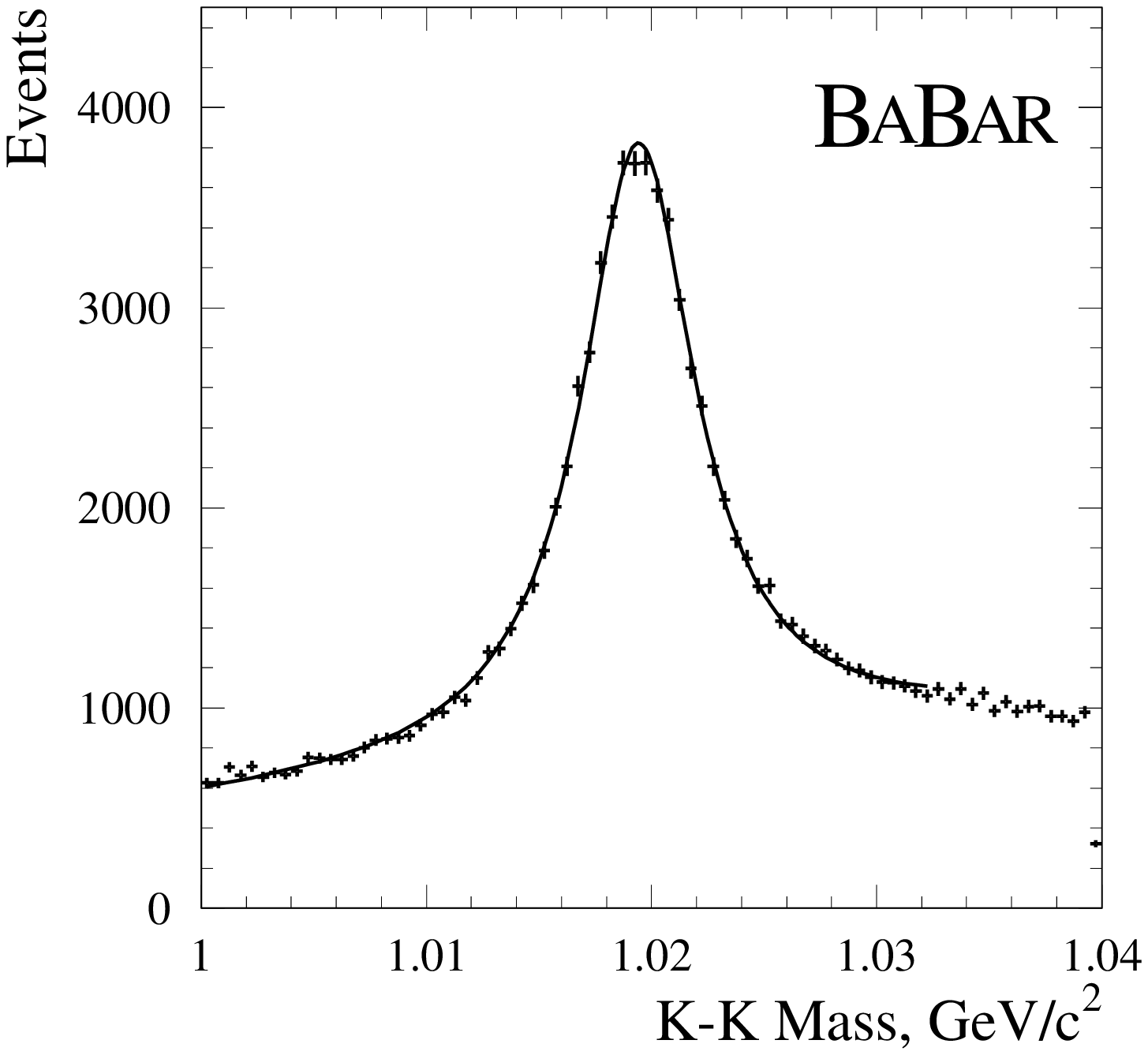}
\caption{\small The $\rm K^+K^-$ invariant mass spectrum for
an integrated luminosity of 1.53 \fbi. The solid line represents a fit using a
Breit-Wigner function and a 1st order polynomial.}
\label{fig:phi}
\end{minipage}
\hfill
\begin{minipage}{.45\twd}
\includegraphics[width=\twd]{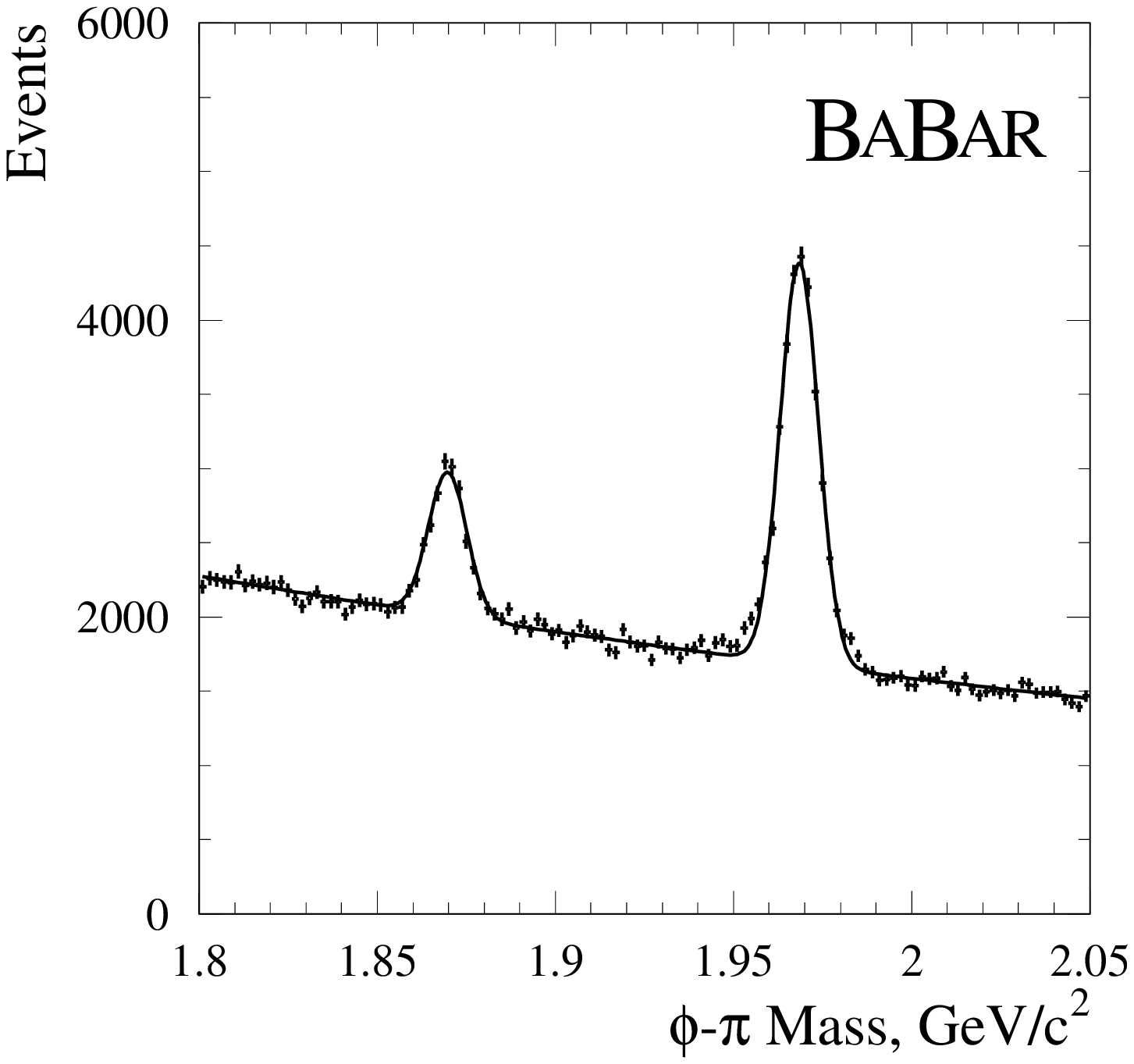}
\caption{\small The $\phi\pi$ invariant mass spectrum for an integrated
luminosity of 7.73 \fbi .}
\label{fig:dstophipi}
\end{minipage}
\end{figure}
\end{center}

Three charged tracks coming from a common vertex 
are then combined to form 
a $\Dspm$ candidate. Two oppositely-charged tracks have to be identified as
kaons, one of these using the basic criteria and the second one using the 
tighter selection. 
The ${\rm K^+K^-}$ 
invariant mass must be within 8\mevcc\ of the nominal $\phi$ mass
(see Fig.~\ref{fig:phi}).
In this particular decay, the $\phi$ meson is polarized 
longitudinally and therefore the angular distribution of the kaons has a 
$\cos^2\theta_{H}$ dependence, where
the $\theta_{H}$ is the angle between the $\rm K^+$ 
in the $\phi$ rest frame and the $\phi$ direction
in the $\Dspm$ rest frame. 
We require $|\cos\theta_{H}|>$0.3, thereby keeping 97.5\% of the 
signal while rejecting about 30\% of the background.

Using the selection criteria described above, a reasonably clean signal of
$\Dspm$ is observed (Fig.~\ref{fig:dstophipi}). The efficiency averaged over
all momenta is (40.5$\pm 1.0$)\%.
It varies as
a function of the $\Dspm$ momentum and ranges from 30\% when the $\Dspm$ is 
at rest in the \ups rest frame to 55\% for $p^* =5$\gevc.
A clear signal of the
Cabibbo-suppressed decay mode \Dphipi \ is also observed. 
A summary of the measured signal properties
is given in the Table~\ref{tab:cuts}.  The final production rates,
however, are obtained from the invariant mass spectra
fitted separately for different momentum intervals 
(Section~\ref{inclusive}).
The measured mass difference m$_{\Dspm }$-m$_{\Dpm }$ agrees  with
the world average value of $99.2 \pm 0.5$\mevcc~\cite{PDG}.

\vspace{-1.5cm}

\begin{table}[htb]
\caption{Fitted parameters for $\Dsphipi$ and
$\Dspmstar \rightarrow \Dspm \gamma$ decay modes.}
\begin{center}
{ \normalsize
\begin{tabular}[angle=90]{|l|c|c|} \hline
 		&\dstophipi & $\Dspmstar \rightarrow \Dspm \gamma$   \\ \hline\hline 

Fit		& $N_{D_S} = 18269 \pm 202$ 		& $N_\Dspmstar = 3029\pm 151$ \\
		& $M = 1968.5 \pm 0.1$\mevcc 	      	& $\Delta M =  143.4\pm 0.3$\mevcc	\\
		& $\sigma = 5.40\pm 0.07$\mevcc  	& $\sigma_{\Delta m} = 7.4\pm 0.4$\mevcc \\ 
		& $M_{D_s^{\pm}}-M_{D^{\pm}} = 98.7\pm 0.2$\mevcc       &		        \\ \hline
\end{tabular}
}
\label{tab:cuts}
\end{center}
\end{table}

\subsection{Inclusive {\boldmath $\Dspm$} momentum spectra}
\label{inclusive}

The number of $\Dspm$ mesons is extracted by fitting the $\phi\pi^{\pm}$ 
invariant mass distribution for 
different momentum ranges in the \ups rest frame. 
The momentum bin width is 200\mevc, which is much larger than
the momentum resolution. 
The $\Dspm$ momentum resolution 
averaged over all momenta obtained from the Monte Carlo 
is $5.6\pm 0.3$\mevc.
The fit function is a single Gaussian distribution, both for the $\Dspm$ and the $\Dpm$.
The width of the Gaussians are constrained to be the same
and the combinatorial background is accounted for by an exponential distribution.
The number of $\Dspm$ in the off-resonance data 
is extracted using the same fit function but with fixed values for M$_\Dpm$, M$_{\Dspm }$ 
and $\sigma$ obtained from the fit to the on-resonance data.

\begin{center}
\begin{figure}[tb]
\begin{minipage}{.48\twd}
\includegraphics[width=\twd]{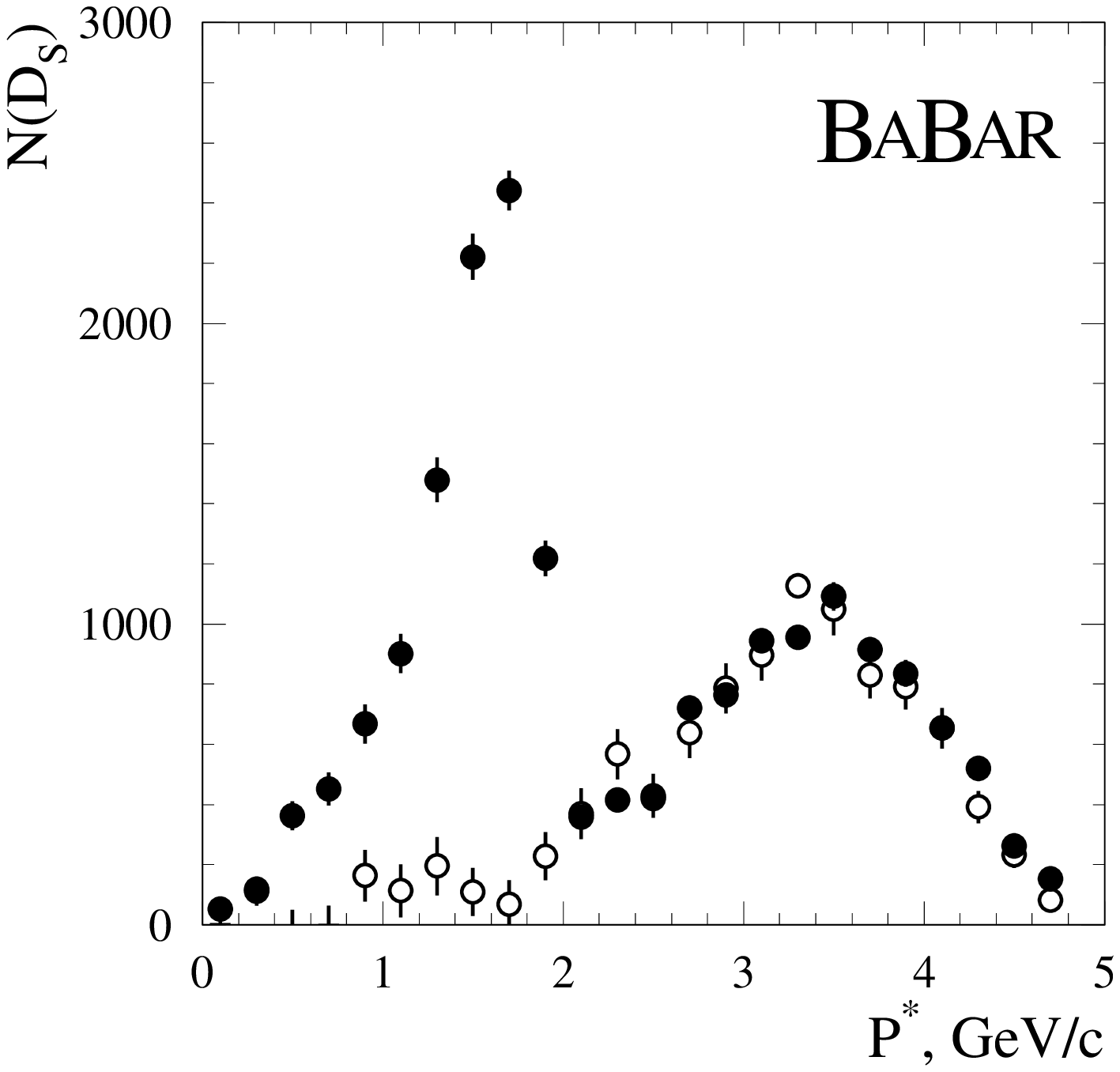}
\caption{\small
The $\Dspm$ momentum spectrum for on-resonance data (solid circles)
 and for scaled off-resonance data (open circles) 
before efficiency correction.}
\label{fig:numds_res}
\end{minipage}
\hfill
\begin{minipage}{.48\twd}
\includegraphics[width=\twd]{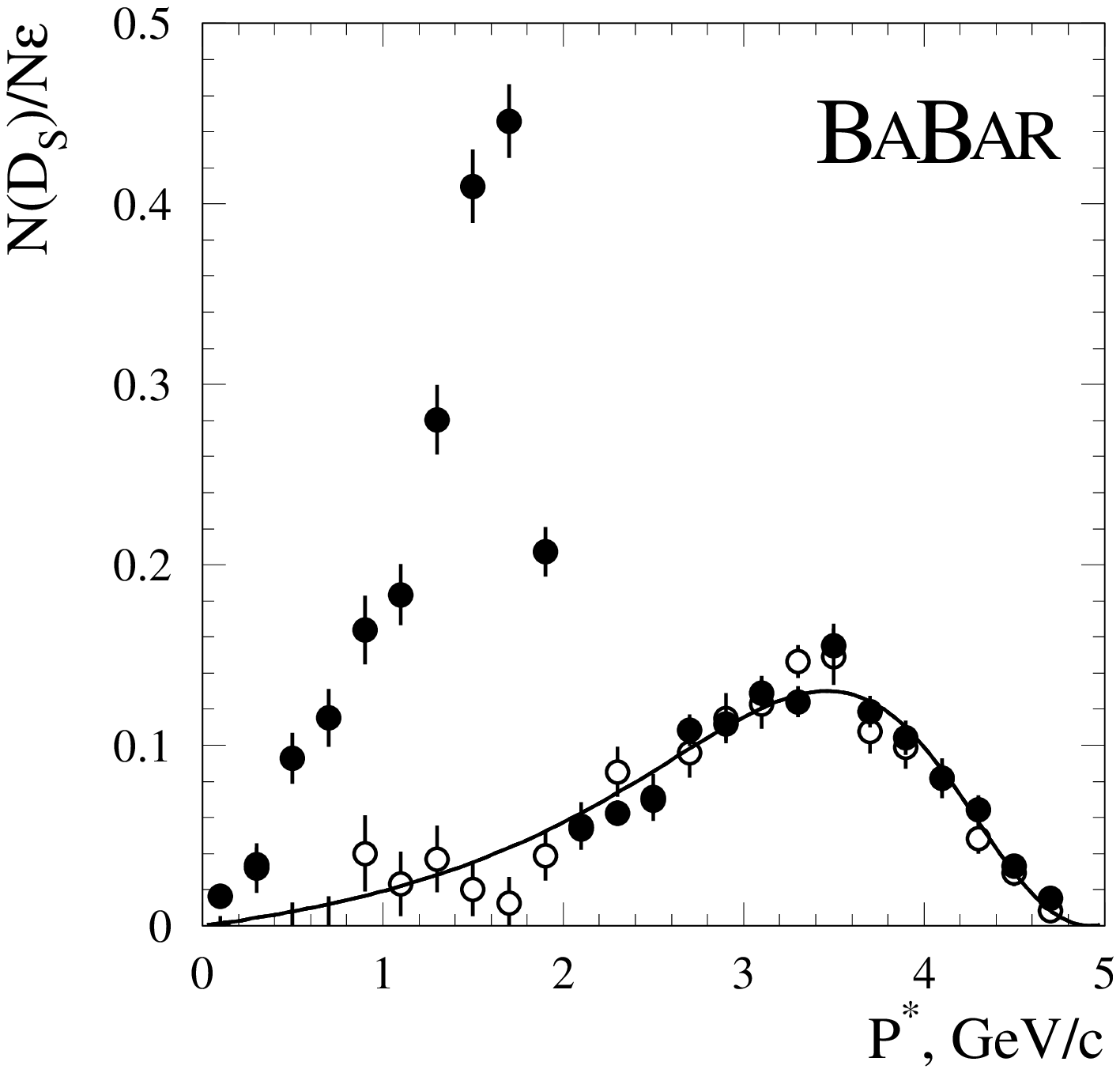}
\caption{\small The on-resonance (solid circles) and scaled off-resonance 
(open circles) D$_S$ momentum spectrum efficiency-corrected. 
The solid line is the result of the fit using Peterson
fragmentation function described in text.}
\label{fig:spec_corr}
\end{minipage}
\end{figure}
\end{center}

The number of reconstructed $\Dspm$ as a function of their momentum 
in the \ups rest frame
is shown in Fig.~\ref{fig:numds_res} for on- and off-resonance data. 
The efficiency-corrected momentum spectrum 
is shown in Fig.~\ref{fig:spec_corr}.

\vspace{-1.5cm}

\begin{table}[htb]
\caption{Analytical expressions for the fragmentation functions.}
\begin{center}
\begin{tabular}[angle=90]{ll} \hline
Name of function 		&  Analytical expression	\\ \hline \hline
Peterson \etal :		& $f(x_p)=\frac{N}{x_p}\left( 1-\frac{1}{x_p}-\frac{\epsilon}{1-x_p} \right)
					 ^{-2}$				\\
Collins and Spiller:		& $f(x_p)=N \left(\frac{1-x_p}{x_p}+\frac{2-x_p}{1-x_p}\epsilon \right)
					(1+x_p^2)\left( 1-\frac{1}{x_p}-\frac{\epsilon}{1-x_p} \right)^{-2}$ \\ 
Kartvelishvili \etal:		& $f(x_p)=Nx_p^\alpha (1-x_p)$		\\ \hline \hline
\end{tabular}
\label{tab:fragmfunc}
\end{center}
\end{table}

In order to determine the $\Dspm$ momentum spectrum from the
continuum, on-resonance data with momentum higher than 2.45\gevc\ 
and off-resonance data scaled according to 
the luminosity ratio have been fitted after efficiency correction
using 3 different 
fragmentation functions (see Table~\ref{tab:fragmfunc}).
The product of branching fraction, ${\mathcal{B}}(\Dsphipi )$, times
cross-section for $\Dspm$ production from continuum, $\sigma (e^+e^-\rightarrow \Dspm X)$, is obtained 
by integrating the function obtained from the fit (Fig.~\ref{fig:spec_corr}). 
The extracted values and $\chi^2$/{\it dof} from the fits are shown in 
Table~\ref{tab:Dsfromccbar}.

Table~\ref{tab:syst1} shows the contribution of the different sources to the total systematic error for
$\sigma (e^+e^-\rightarrow \Dspm X)\cdot{\mathcal{B}}(\Dsphipi )$.
Using the best fit, which is obtained with the Peterson function, we find
$\sigma (e^+e^-\rightarrow \Dspm X)\cdot{\mathcal{B}}(\Dsphipi )$ 
= 8.29$\pm$0.41$\pm$0.69~pb.
From a comparison of the results obtained using the other
two parameterizations, we assign a conservative systematic error of 2\%
due to the assumed functional form. 

The measured values are in good agreement with previously
published results \cite{argus}.
The momentum spectrum of the D$_s$ produced in B decays is obtained 
by subtracting bin-by-bin  
the value of the fit function to the on-resonance data 
after efficiency correction
(Fig.~\ref{fig:spec_corr_sub}).

\begin{table}[htb]
\caption{The parameters for the different fragmentation functions obtained
from the fit and the measured cross section 
$\sigma (e^+e^-\rightarrow \Dspm X)\cdot{\mathcal{B}}(\Dsphipi )$,
and the $\chi^2$/{\it dof} of the fit.  Only the statistical errors
are given.}\vspace{0.3cm}
\begin{center}
\begin{tabular}[angle=90]{lccc} \hline
Name of function 		&  Shape parameter	& $\sigma (e^+e^-\rightarrow \Dspm X)\cdot{\mathcal{B}}(\Dsphipi )$, pb	& $\chi^2$/{\it dof}  \\ \hline \hline
Peterson \etal :		& $\epsilon$=(12.5$\pm$0.6) $\times$ 10$^{-2}$ 	& 8.29$\pm$0.41		 	& 1.286\\
Collins and Spiller:		& $\epsilon$=(37.6$\pm$2.8) $\times$ 10$^{-2}$	& 8.69$\pm$0.46			& 3.559\\ 
Kartvelishvili \etal:		& $\alpha$=1.91$\pm$0.07			& 8.63$\pm$0.33			& 5.338\\ \hline \hline
\end{tabular}
\label{tab:Dsfromccbar}
\end{center}
\end{table}

\begin{table}[htb]
\caption{The systematic errors for 
$\sigma (e^+e^-\rightarrow D^{\pm}_sX)\cdot{\mathcal{B}}(\Dsphipi )$.}
\begin{center}
\begin{tabular}[angle=90]{lr} \hline \hline
Source	& Error (\%) \\ \hline 
$\mathcal{B}(\phi\rightarrow {\rm K^+K^-})$	&	1.6   \\  
Particle id  efficiency				&       0.8   \\ 
Tracking efficiency				&       7.5    \\ 
Luminosity					&	3.0 	\\ \hline
Total systematic error 				&	8.3 	\\ \hline \hline 
\end{tabular}
\label{tab:syst1}
\end{center}
\end{table}

\subsection{Inclusive {\boldmath $\Dspm$} branching
fraction in {\boldmath $B$} decays}

By integrating the efficiency corrected momentum distribution, a total 
D$_s$ yield from B meson decays of $37050\pm 950$ events is found.
This corresponds to the inclusive branching fraction of
\be
{\mathcal{B}}(B\rightarrow \Dspm X) = \Biggl[(11.90\pm0.30\pm1.07)\times 
\frac{3.6\pm0.9\%}{{\mathcal{B}}(\Dsphipi )}\Biggr]\%,
\ee
where the first error is statistical, the second is systematic 
and the third is the contribution 
of the $\Dsphipi$~branching fraction uncertainty \cite{PDG}. 
Recognizing that this last uncertainty is common to all measurements,
our result is slightly higher than the world average 
(10.0$\pm 0.6$ \cite{PDG}) and in good agreement 
with the most precise measurement performed by CLEO \cite{cleo:dsinc}.
The different sources  of systematic errors 
are given in detail in Table~\ref{tab:syst2}. 
The dominant uncertainty comes from knowledge of the tracking efficiency,
which is still the subject of detailed study \cite{babdet}.

As a cross check of the continuum subtraction procedure, 
we also subtracted directly the off-resonance data 
scaled by the luminosity 
ratio for on- and off-resonance. 
By this means, one obtains an inclusive branching fraction
${\mathcal{B}}(B\rightarrow \Dspm X) = 12.0\pm 0.5\pm 1.1$\%,  
in agreement with the value reported above.

\begin{center}
\begin{table}[htb]
\begin{minipage}{.48\twd}
\caption{Systematic errors for ${\mathcal{B}}(B\rightarrow \Dspm X)$.}\vspace{0.3cm}
\begin{center}
\begin{tabular}[angle=90]{lr} \hline \hline
Source	& Error (\%) \\ \hline 
Signal shape 					&	0.9	\\
Background shape 				&	0.4	\\
Continuum subtraction				&	1.8	\\
Monte Carlo statistics				&	2.0	\\
Bin width					&       0.7     \\ \hline
Total for D$_S$ yield				&	2.9	\\
N$_{B\overline{B}}$				&	3.6	\\
$\mathcal{B}(\phi\rightarrow {\rm K^+K^-})$	&	1.6   \\  
Particle id  efficiency				&       0.8   \\ 
Tracking efficiency				&       7.5    \\ \hline
Total systematic error 				&	9.0 	\\ \hline \hline 
\end{tabular}
\label{tab:syst2}
\end{center}
\end{minipage}
\hfill
\begin{minipage}{.48\twd}
\caption{Systematic errors for ${\mathcal{B}}(B\rightarrow \Dspmstar X)$.}
\begin{center}
\begin{tabular}[angle=90]{lr} \hline \hline
Source	& Error (\%) \\ \hline 
Signal shape 					&	5.0	\\
Continuum subtraction				&	1.2	\\
Monte Carlo statistics				&	4.8	\\
Bin width					&       3.0     \\ \hline
Total for D$_S$ yield				&	7.7	\\
N$_{B\overline{B}}$				&	3.6	\\
$\mathcal{B}(D_S^{*+}\rightarrow D_S\gamma)$    &	2.7	\\
Photon efficiency				&       2.5     \\
$\mathcal{B}(\phi\rightarrow {\rm K^+K^-})$	&	1.6   \\  
Particle id  efficiency				&       0.8   \\ 
Tracking efficiency				&       7.5    \\ \hline
Total systematic error 				&	12.0 	\\ \hline \hline 
\end{tabular}
\label{tab:syst3}
\end{center}
\end{minipage}
\end{table}
\end{center}

\section{Inclusive {\boldmath $\Dspmstar$} production}

\subsection{{\boldmath $\Dspmstar$} reconstruction}

\begin{center}
\begin{figure}[htb]
\begin{minipage}{.48\twd}
\begin{center}
{\Large \bf Preliminary}\\
\includegraphics[width=\twd]{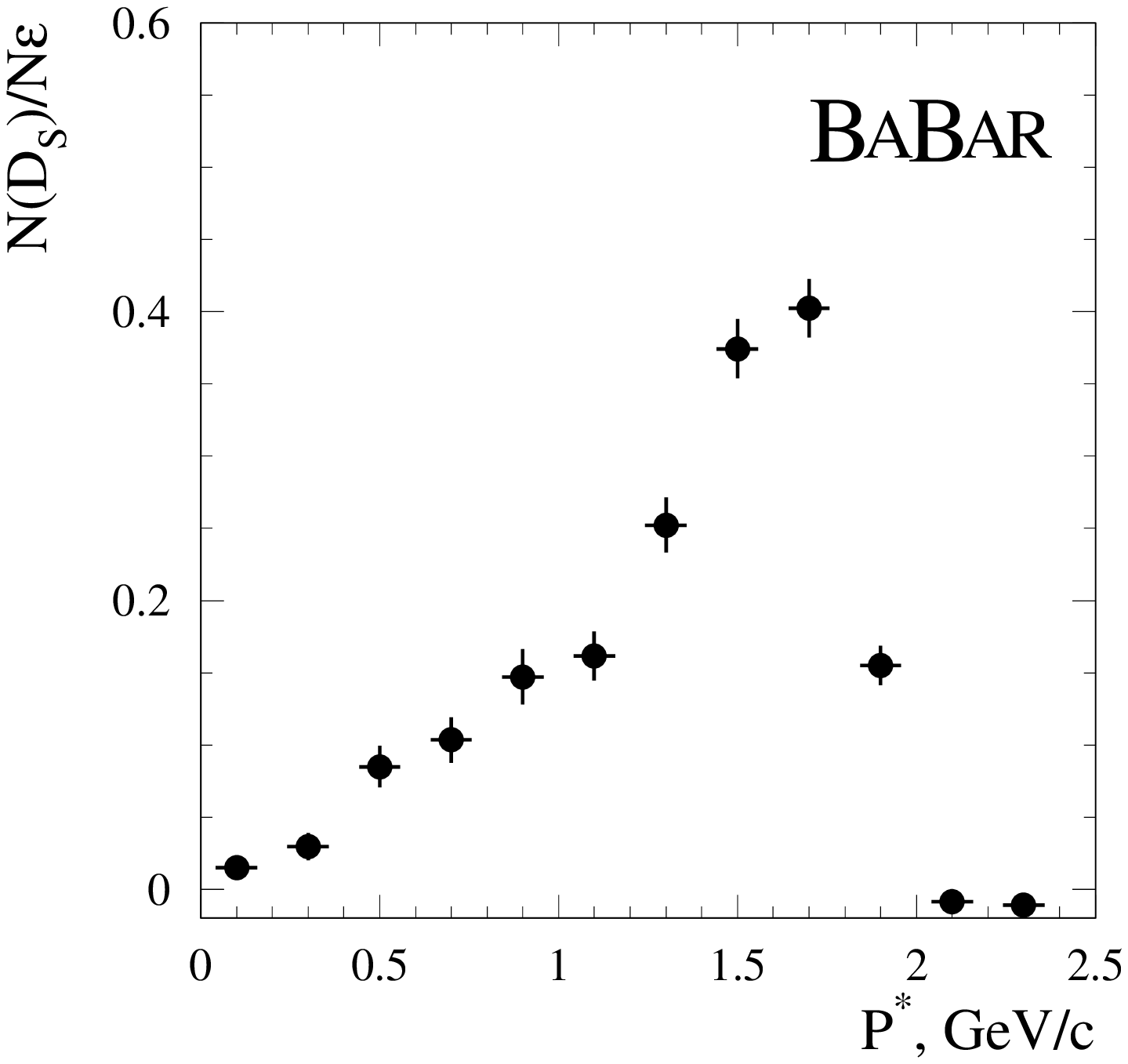}
\end{center}
\caption{\small
The D$_s$ momentum spectrum efficiency-corrected after subtraction 
of the value of the
fitted curve. The Peterson fragmentation function was used for the fit of the continuum.}
\label{fig:spec_corr_sub}
\end{minipage}
\hfill
\begin{minipage}{.48\twd}
\includegraphics[width=\twd]{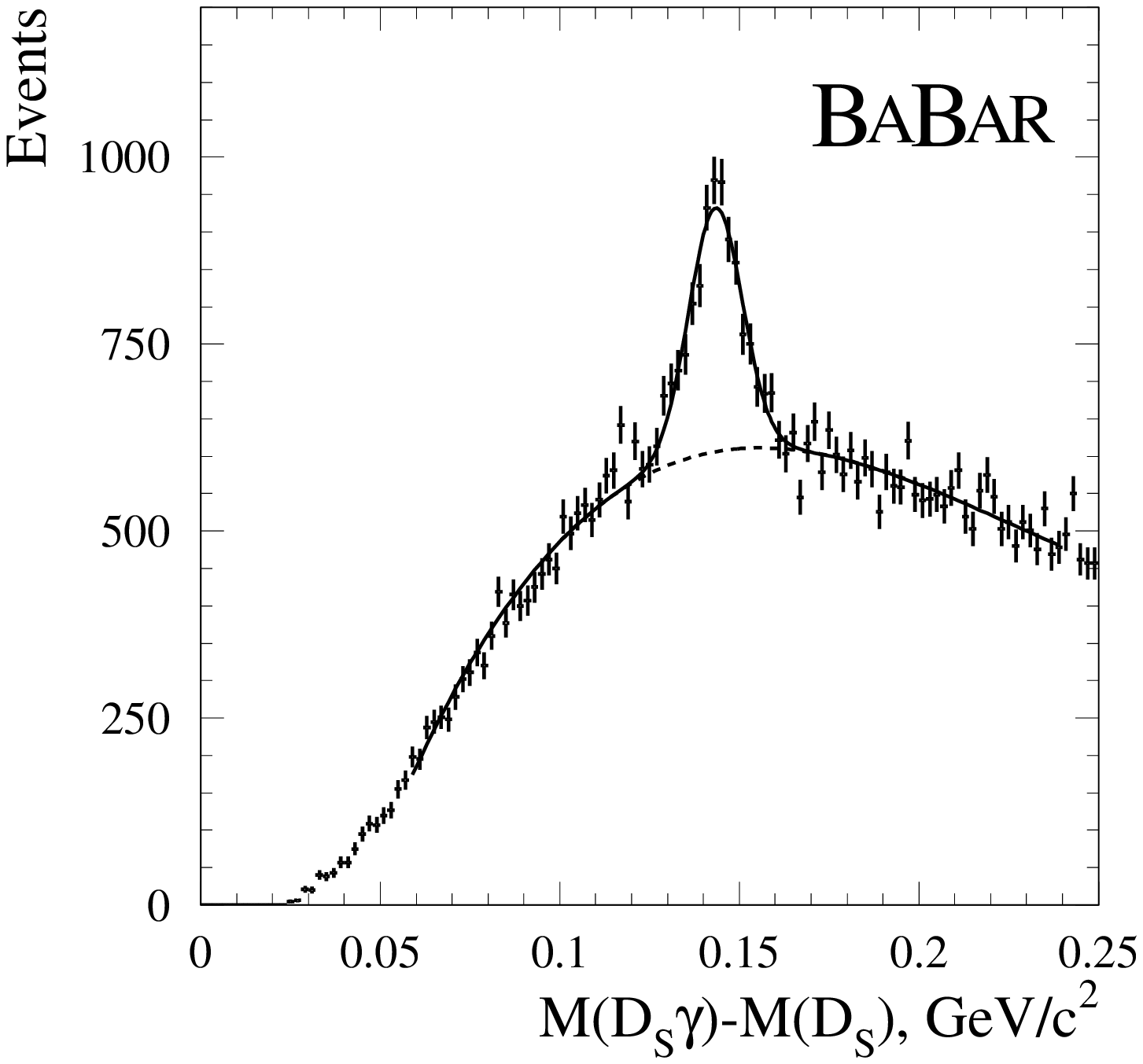}
\caption{\small
$\Delta M = M_{{\rm D_s\gamma}} -  M_{{\rm D_s}}$ mass spectrum 
for an integrated luminosity of 7.73 \fbi. 
The fit function is a single Gaussian for the signal and 
a third-order polynomial for the background.}
\label{fig:dsstar}
\end{minipage}
\end{figure}
\end{center}

$\Dspmstar$ mesons are reconstructed using the decay 
$\Dspmstar \rightarrow \Dspm \gamma$ with the subsequent decay
$\Dspm \rightarrow \phi\pi$.
$\Dspm$ candidates are selected by requiring the $\phi\pi$ invariant mass to
be within 2.5 standard deviations ($\sigma$) of the peak value. These are then combined 
with ``single photons'' from the event. The later are defined 
by the following criteria:

\begin{itemize}
\item $E_\gamma > 50$\mev\ where $E_{\gamma}$ 
is the photon energy in the laboratory frame
\item $E_\gamma^* > 110$\mev\ where $E^*_{\gamma}$ 
is the photon energy in the \ups frame
\item In order to reduce the combinatoric background,
the candidate photon should not form a $\pi^0$ with 
$E_{\gamma\gamma}^* > 200$\mev\
when combined with any other photon in the event. The $\pi^0$ mass window
is $115 < M_{\gamma\gamma} < 155$\mevcc.
\end{itemize}
The distribution of the mass difference 
$\Delta M = M_{\Dspm \gamma} -  M_{\Dspm}$ 
is shown in the Fig.~\ref{fig:dsstar}. 
A clear peak with $3030\pm 150$ events is observed. The parameters obtained 
from the fit are summarized in Table~\ref{tab:cuts}.

\subsection{Inclusive {\boldmath $\Dspmstar$} momentum spectra}

The decay $\rm D_s^{*\pm}\rightarrow D_s^{\pm}\gamma$, $\Dsphipi$ is used 
for the measurement of the $\Dspmstar$ inclusive branching fraction
and the momentum spectrum. 
The number of $\Dspmstar$ mesons is extracted by fitting the $\Delta M = M_{{\rm D_s\gamma}}-M_{{\rm D_s}}$ 
invariant mass distribution for  the 
different momentum ranges in the \ups rest frame. 
A momentum bin width of 400\mevc\ was chosen.

\begin{center}
\begin{figure}
\begin{minipage}{.48\twd}
\includegraphics[width=\twd]{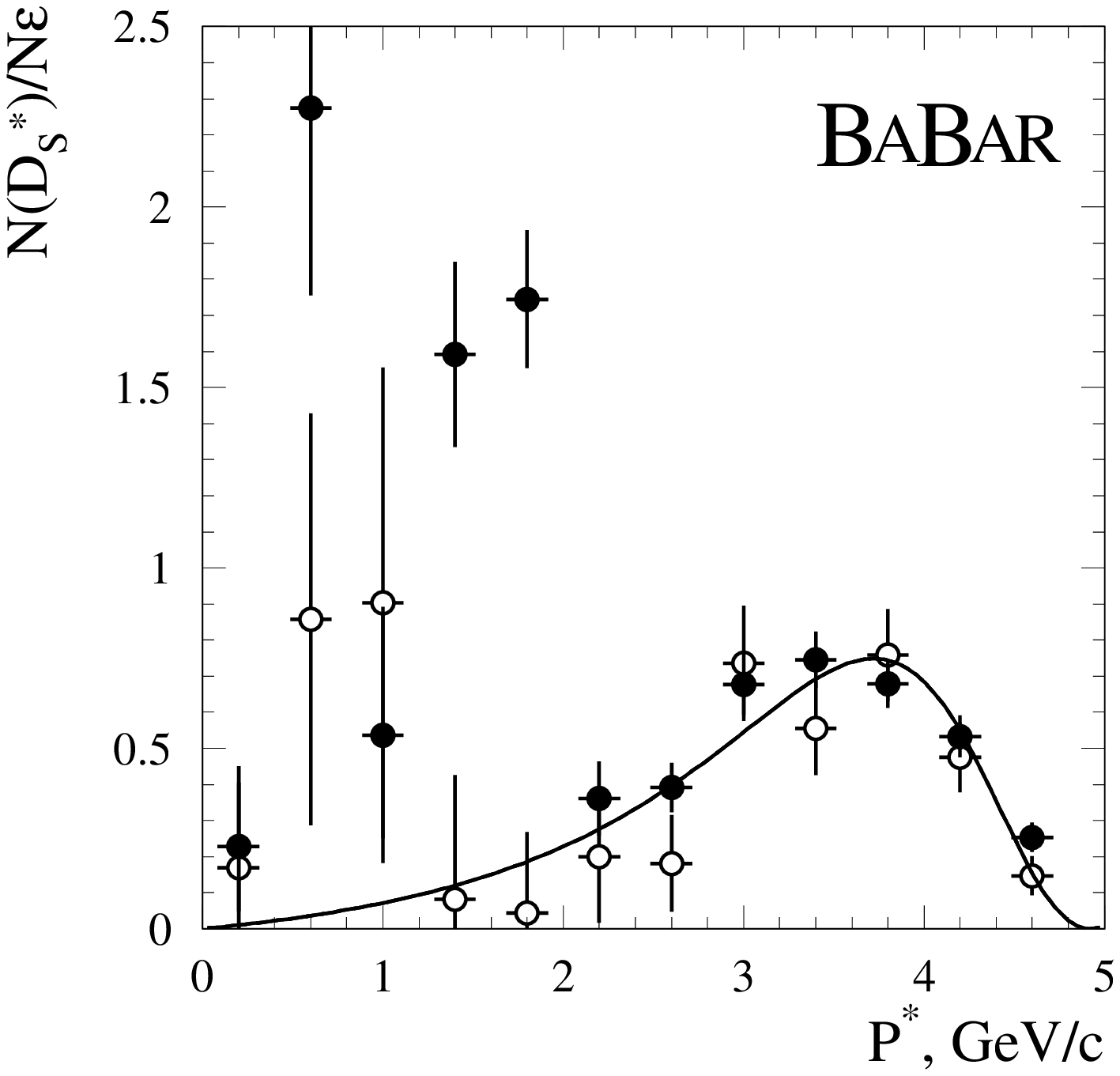}
\caption{\small
The on-resonance (solid circles) and scaled 
off-resonance (open circles) $\Dspmstar$ momentum 
spectrum after efficiency correction. 
The solid line shows the fit using the Peterson
fragmentation.}
\label{fig:numdsst_corr}
\end{minipage}
\hfill
\begin{minipage}{.48\twd}
\begin{center}
{\Large \bf Preliminary}\\
\end{center}
\includegraphics[width=\twd]{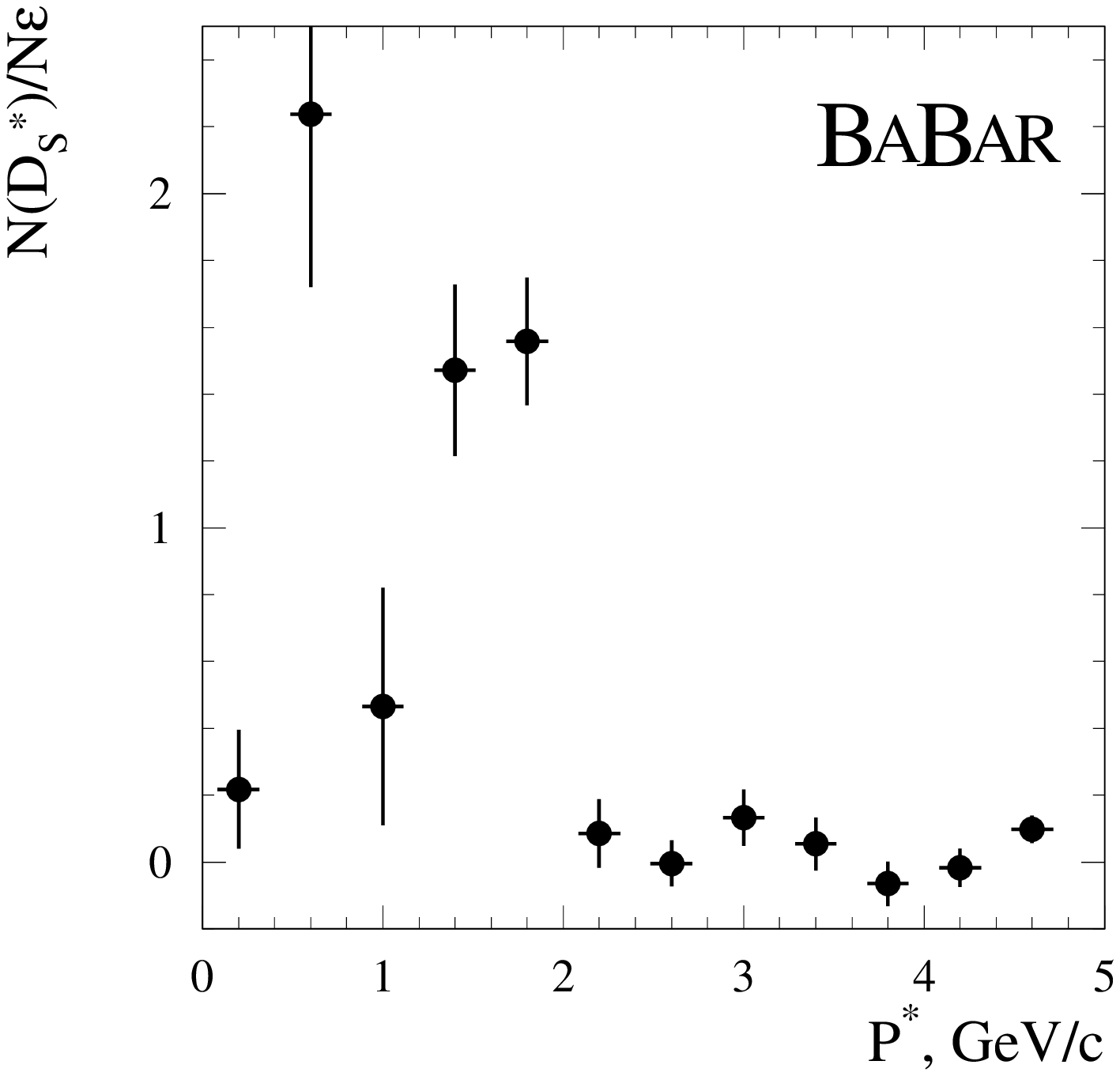}
\caption{\small
The $\Dspmstar$ spectrum after efficiency correction 
and continuum subtraction using
the result of the fit. The Peterson fragmentation function is used for 
the fit of the continuum.}
\label{fig:dsst_spec_corr_sub}
\end{minipage}
\end{figure}
\end{center}

The efficiency corrected momentum spectrum is shown in Fig.~\ref{fig:numdsst_corr}. Both on-
and off-resonance points corresponding to $\Dspmstar$ mesons produced 
from the continuum have been 
fit using different fragmentation functions (Table~\ref{tab:fragmfunc}). 
The cross section for $\Dspmstar$ produced from continuum 
and the values of the fit parameters are shown in 
Table~\ref{tab:Dsstfromccbar}.

Fig.~\ref{fig:dsst_spec_corr_sub} shows the momentum spectrum of 
$\Dspmstar$ produced in B decays where the Peterson fragmentation function 
is used for continuum extrapolation. Using this distribution, we find
for the continuum cross section
$\sigma (e^+e^-\rightarrow \Dspmstar X)\cdot{\mathcal{B}}(\Dsphipi )$ 
= 3.48$\pm$0.39$\pm$0.38~pb.

\vspace{-0.5cm}

\begin{table}
\caption{The parameters for the different fragmentation functions, the measured cross section 
$\sigma (e^+e^-\rightarrow \Dspmstar X)\cdot{\mathcal{B}}(\Dsphipi )$,
and the $\chi^2$/{\it dof} obtained from the fit.  Only the statistical errors
are given.}\vspace{0.3cm}
\begin{center}
\begin{tabular}[angle=90]{lccc} \hline
Name of function 		&  Parameter	& $\sigma (e^+e^-\rightarrow D^{*\pm}_sX)\cdot{\mathcal{B}}(D^+_S\rightarrow \phi\pi^+)$, pb	& 
$\chi^2$/{\it dof}  \\ \hline \hline
Peterson \etal :		& $\epsilon$=(7.9$\pm$0.8)$\times$ 10$^{-2}$ 	& 3.48$\pm$0.39 	& 1.260\\
Collins and Spiller:		& $\epsilon$=(19.3$\pm$2.3)$\times$ 10$^{-2}$	& 3.75$\pm$0.42		& 1.288\\ 
Kartvelishvili \etal:		& $\alpha$=2.6$\pm$0.2				& 3.61$\pm$0.29		& 1.725\\ \hline \hline
\end{tabular}
\label{tab:Dsstfromccbar}
\end{center}
\end{table}

\subsection{Inclusive {\boldmath $\Dspmstar$} branching
fraction in {\boldmath $B$} decays}

In the same way as for the $\Dspm$ result, we integrate the 
efficiency corrected 
$\Dspmstar$ distribution and obtain a total yield from B meson 
decays of $19300\pm 1900$ events.  From this we find
the inclusive branching fraction to be
\be
{\mathcal{B}}(B\rightarrow \Dspmstar X) = \Biggl[(6.8\pm0.7\pm0.8)\times 
\frac{3.6\pm0.9\%}{{\mathcal{B}}(\Dsphipi )}\Biggr]\%,
\ee
where the systematic errors are given in detail in Table~\ref{tab:syst3}.

\section{Branching fraction for {\boldmath \BDstarDss} decays}

In addition to the measurements of inclusive production rates
for $\Dspm$ and $\Dspmstar$, we have extracted the branching ratios for the decays
\BDstarDs \ and \BDstarDsstar \ based on a partial reconstruction
method.

\subsection{The partial reconstruction method}

As discussed in the introduction, no attempt is made to reconstruct the 
$\DN$ decays. One combines a pion with the reconstructed $\Dspms$ where
the total $\Dspms -\pi$ charge is zero and, assuming that their origin 
is a $\BN$ meson, we
calculate the missing invariant mass. This should be the $\DN$ mass if
the hypothesis is correct\footnote{All calculations in this section are performed in the \ups rest frame.}.
Without the constraint of the $\DN$ mass, the direction of
the B meson is unknown. Although its angle with respect to $\Dspm$ 
direction can be deduced, 
the angle $\phi$ around this direction is undetermined.
Using the beam energy constraint,
the missing mass, which still depending on the unknown angle $\phi$ of the
$B^0$ momentum vector, is computed from:
\begin{equation}
m_{\rm miss} = \sqrt { (E_{\rm beam} - E_{\Dspm }  - E_{\pi} )^2 
- ( \vec{p}_B - \vec{p}_{\Dspm }- \vec{p}_{\pi})^2  }. 
\end{equation}
In this analysis the missing mass is defined 
using an arbitrary choice for the angle $\phi$.
We use the convention
that the direction of the $\BN$ meson lies
in the plane $\{ \vec{p}_{\pi},\vec{p}_{\Dspms} \}$.

\subsection{Signal extraction}

Fully reconstructed $\Dspm$ and $\Dspmstar$ are selected by requiring
the measured $\phi\pi^\pm$ mass or 
$\Delta m=m_{\phi\pi^\pm\gamma}-m_{\phi\pi^\pm}$ to be within 2.5 
$\sigma$ of the fitted mean value.
Because of high combinatorial background in the mode with a $\Dspmstar$, 
one may find several $\Dspmstar$ candidates in an event.
Therefore, we form a $\chi^2$ for each candidate defined by:
\be
\chi^2=\biggl(\frac{M_{\phi}^{rec}-M_{\phi}^{mean}}{\sigma_{\phi}}\biggr)^2+
\biggl(\frac{M_{D_s}^{rec}-M_{D_s}^{mean}}{\sigma_{D_s}}\biggr)^2+
\biggl(\frac{M_{\Delta m}^{rec}-M_{\Delta m}^{mean}}{\sigma_{\Delta m}}\biggr)^2 \;,
\ee
and take the candidate with the lowest value.
The $\Dspms-\pi$ pairs satisfying the kinematic constraints 
for the decay \BDstarDss \ are fitted to a common vertex.
To reduce further the continuum background, we use 
the event shape variable $R_2$, defined as 
the ratio of the second to zeroth order Fox-Wolfram moment,
and require $R_2 < 0.35$.

\begin{center}
\begin{figure}[hbt]
\begin{minipage}{.48\twd}
\includegraphics[width=\twd]{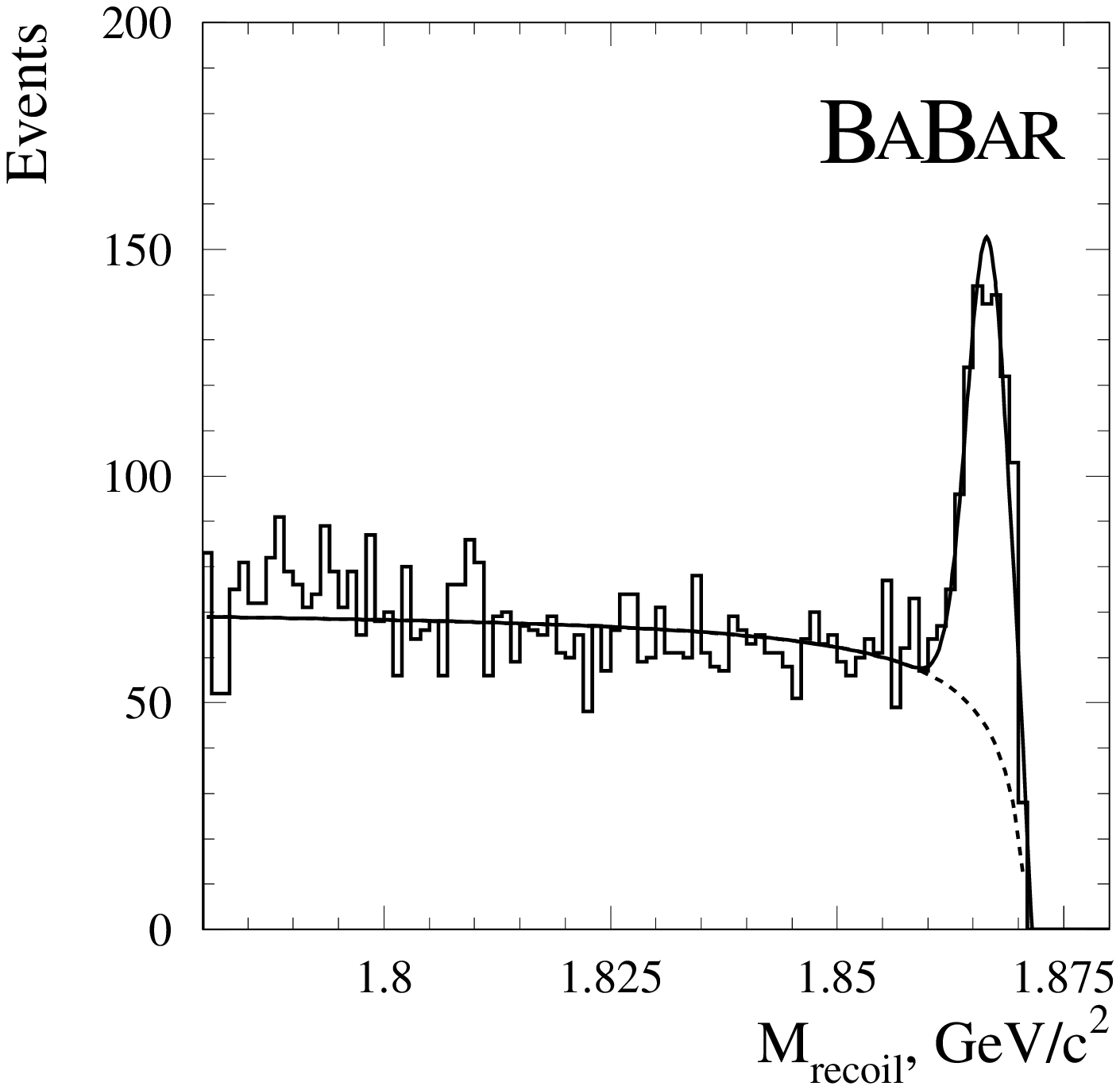}
\caption{\small
The missing mass distribution for the $\Dspm$-$\pi$ system.  
The solid line shows the result of the fit 
using the function described in the text.}
\label{fig:dsdst}
\end{minipage}
\hfill
\begin{minipage}{.48\twd}
\includegraphics[width=\twd]{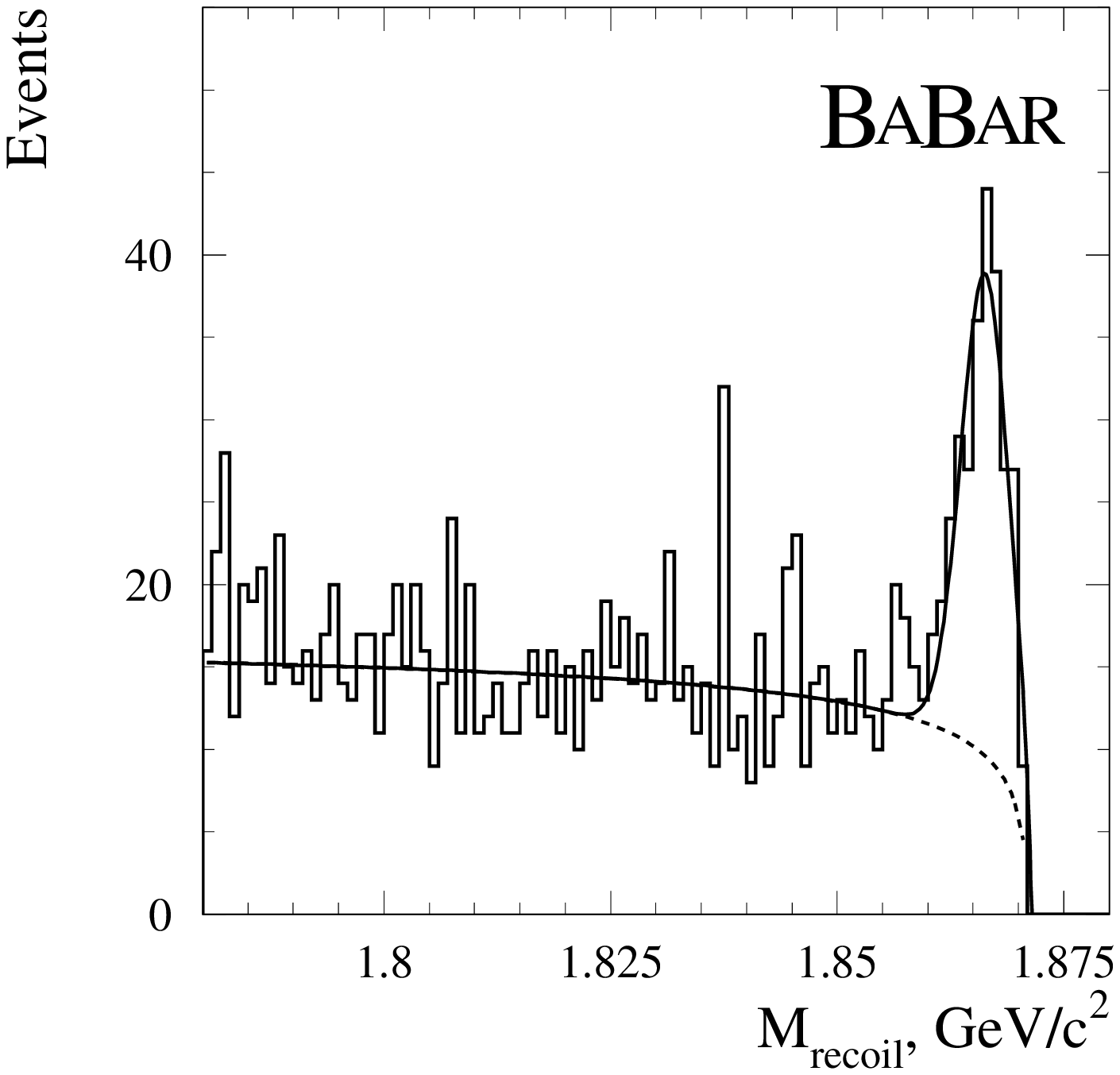}
\caption{\small
The missing mass distribution for the $\Dspmstar$-$\pi$ system.  
The solid line shows the result of the fit 
using the function described in the text.}
\label{fig:dsstdst}
\end{minipage}
\end{figure}
\end{center}

The missing mass distributions for the 
$\Dspm-\pi$ and $\Dspmstar-\pi$ are shown in Fig.~\ref{fig:dsdst} 
and~\ref{fig:dsstdst} respectively. A clear signal is observed for both decays.
The missing mass distribution is fitted with  
the sum of a Gaussian distribution for the signal 
and a background function given by
\be
f_B(x) = \frac{C_1\biggl(x_0-x\biggr)^{C_2}}{C_3+\biggl(x_0-x\biggr)^{C_2}}
\;,
\ee
where $x$ is the calculated missing mass,
$C_i$ are the parameters of the fit and $x_0$ is the end point,
${\rm m}_{\rm D^*}-{\rm m}_{\pi} = 1.871$\gevcc.
The results of the fits for both decay modes are summarized in 
Table~\ref{tab:inc}.

\begin{table}[htb]
\caption{Selection criteria and fit parameters for the missing mass 
distribution in partially reconstructed \BDstarDss decays.}\vspace{0.3cm}
\begin{center}
\begin{tabular}[angle=90]{|l|l|} \hline
{ $B\rightarrow {\rm D}_s^{+}{\rm D}^{*-}$} &{$B\rightarrow {\rm D}_s^{*+}{\rm D}^{*-}$} \\ \hline \hline 


$\chi^2/{\it dof}$=1.25		& $\chi^2/{\it dof}$=	1.09		\\
$N_{\rm ev}$ = 628$\pm$55 events    & $N_{\rm ev} = 195\pm 29$ events	\\
$m_{\rm miss} = 1866.7\pm 0.2$\mevcc & $m_{\rm miss} = 1866.3\pm 0.2$\mevcc  \\
$\sigma = 2.31\pm 0.15$\mevcc	   	& $\sigma = 2.66\pm 0.36$\mevcc \\ \hline
\end{tabular}
\label{tab:inc}
\end{center}
\end{table}

\subsection{Branching fractions for \boldmath \BDstarDs\ and \BDstarDsstar}

A Monte Carlo simulation of the \BDstarDss decay modes 
has been used to find the efficiencies. It is important to note that
the \BDstarDsstar\ decay mode contributes 
to the missing mass distribution for the $\Dspm -\pi$ system,
even though there is a missing photon from the $\Dspmstar$.
We show in Table~\ref{tab:eff_inc} the reconstruction efficiencies 
for the different modes.

\begin{table}[htb]
\caption{The efficiencies for the partially reconstructed \BDstarDss 
decay modes. 
The columns show the contribution of the different generated modes 
to the $\Dspm-\pi$ and $\Dspmstar-\pi$ missing mass distributions
in the signal region.
}\vspace{0.3cm}
\begin{center}
\begin{tabular}[angle=90]{|l|c|c|} \hline
& \multicolumn{2}{|c|}{Reconstructed mode} \\ \cline{2-3}
True mode				&  $\Dspm-\pi$		& $\Dspmstar-\pi$   	\\ \hline\hline	
\BDstarDs			&  32.8$\pm$1.8\%	&			\\
\BDstarDsstar long.pol.		&  15.8$\pm$1.2\%	& 9.1 $\pm$0.9\%	\\
\BDstarDsstar transv.pol.	&  14.2$\pm$1.1\%	& 6.0 $\pm$0.7\%	\\ \hline	
\end{tabular}
\label{tab:eff_inc}
\end{center}
\end{table}
Although the amount of feed through from \BDstarDsstar\ to \BDstarDs\
depends on the $\Dspmstar$ polarization, one sees from 
Table~\ref{tab:eff_inc} that the variation is small.
However the reconstruction efficiency for 
\BDstarDsstar \ has a much larger dependence on the polarization. 
Since this polarization is not known, we use the average efficiency of 
($7.5 \pm 1.5$)\%, where the systematic error of $1.5\%$ is
derived by comparing the efficiencies from the two
polarization states.  This is combined
with the other systematic errors which are in common with the inclusive
branching fractions presented in the previous sections.


For the measurement of the  \BDstarDsstar\ branching fraction, 
the contribution to the missing mass peak from \BDstarDs, where a 
random $\gamma$ is associated to the $\Dspm$, is negligible.
The contribution of \BDstarDsstar\ to the mode \BDstarDs\
is then subtracted to determine the branching fraction for the latter.
The results are given
in Table~\ref{tab:br}.  The first error is statistical, the third 
reflects the uncertainty due to the error in the branching ratio 
for $\Dspm \rightarrow \phi \pi^{\pm}$, and the second error represents
all remaining systematics.  This last is dominated by the uncertainty due
to the dependence of the efficiency on the polarization of the final
state.
 
\begin{table}[htb]
\caption{The measured branching fraction for \BDstarDs\ and \BDstarDsstar .}
\begin{center}
{\Large \bf Preliminary}\\
\begin{tabular}[angle=90]{|l|l|} \hline
{ $B\rightarrow {\rm D}_s^{+}{\rm D}^{*-}$}		&{$B\rightarrow {\rm D}_s^{*+}{\rm D}^{*-}$} \\ \hline \hline 
$\mathcal{B}=$ (7.1$\pm$2.4$\pm$2.5$\pm$1.8) $\times$ 10$^{-3}$ & $\mathcal{B}$=2.54$\pm$0.38$\pm$0.53$\pm$0.64\% \\ 
PDG: $\mathcal{B}$=(9.6$\pm$3.4) $\times$ 10$^{-3}$  & PDG: $\mathcal{B}$=2.0$\pm$0.7\% \\ \hline
\end{tabular}
\label{tab:br}
\end{center}
\end{table}

Finally, one should note that the reconstructed \BDstarDsstar\
events should allow us to measure the polarization of the $\Dspmstar$ in these decays and
therefore, in future analyses, it will be possible to 
reduce the systematic error from this source.

\subsection{Background cross checks}

\begin{center}
\begin{figure}[htb]
\begin{minipage}{.48\twd}
\includegraphics[width=\twd]{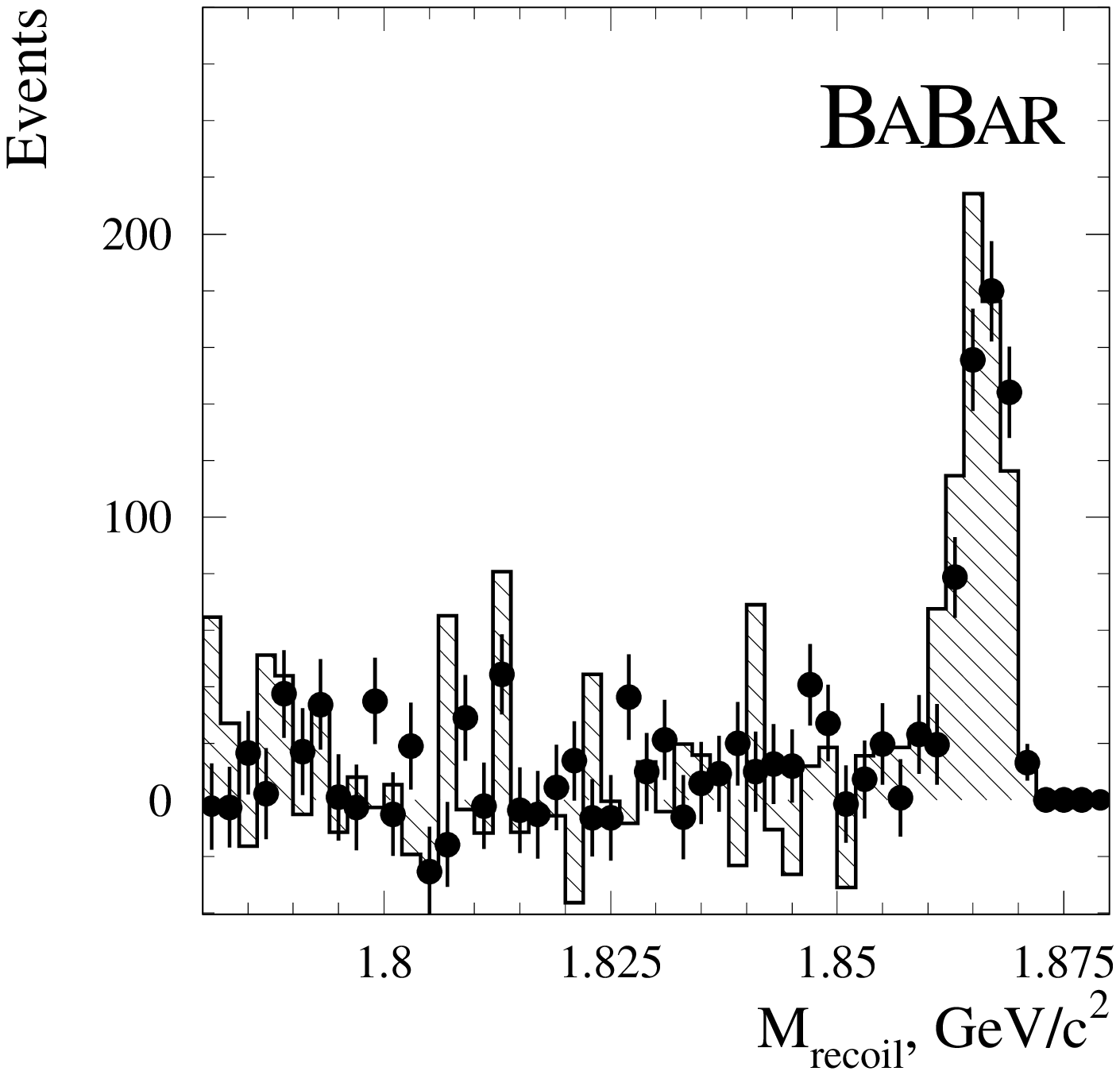}
\caption{\small
The missing mass distribution for the $\Dspm$-$\pi$ system from 
data (points) and Monte Carlo (histogram)
after background subtraction (see text).}
\label{fig:dsdst_mc_sub}
\end{minipage}
\hfill
\begin{minipage}{.48\twd}
\includegraphics[width=\twd]{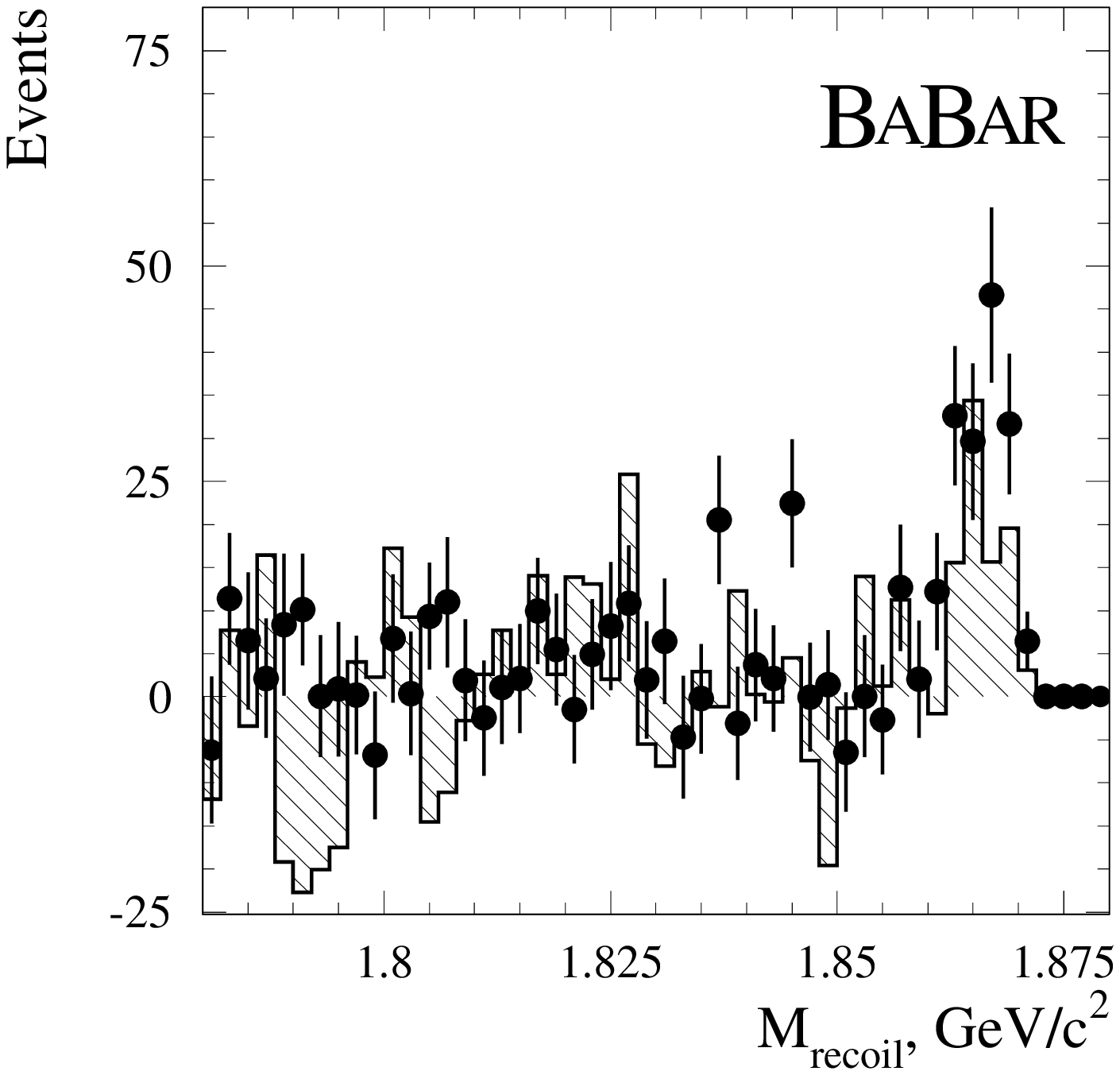}
\caption{\small
The missing mass distribution for $\Dspmstar$-$\pi$ system from 
data (points) and Monte Carlo (histogram)
after background subtraction (see text).}
\label{fig:dsstdst_mc_sub}
\end{minipage}
\end{figure}
\end{center}

In order to investigate further the shape of the background which is subtracted
for estimating the signal, we have compared the Monte Carlo to the data. 
Several types of backgrounds contribute in the signal region:
\begin{enumerate}
\item 
Fake $\Dspms$ and a random pion (for example coming from the other B).
\item 
Fake $\Dspms$ and correlated pion (for example coming from the same B).
\item
True $\Dspms$ and a random pion.
\item 
True $\Dspms$ and a correlated pion.
\end{enumerate}
Table~\ref{tab:bgr} shows the different types 
of backgrounds and the methods which are used to determine their level.  
Background types {\bf 1 + \bf 3} 
are obtained by flipping the $\Dspms$ direction. 
Background types {\bf 1 + \bf 2}
are extracted using the sidebands of the $\Dspms$ mass distribution. 
For this purpose, we take 1.89$<$ M$_{\Dspm}<$1.95 and 1.985$<$ M$_{\Dspm}<$2.05~GeV$/c^2$ 
for the $\Dspm$-$\pi$ system, and $\Delta$M$_\Dspmstar$ 
170$<\Delta$M$_\Dspmstar <$300~MeV$/c^2$ for $\Dspmstar$-$\pi$. 
By flipping the $\Dspms$ direction for the sidebands we find the contribution 
of background type {\bf 1}. 
Therefore the difference between the distributions 
for flipped and non-flipped $\Dspms$ direction for the sidebands gives the type {\bf 2} background contribution and thus
it is possible to find the contribution of background types 
{\bf 1 + \bf 2 + \bf 3} from data alone.
Fig.~\ref{fig:dsdst_mc_sub} and~\ref{fig:dsstdst_mc_sub} show the resulting signal 
after their subtraction. 
The remaining background component is quite small and is estimated from the Monte Carlo. 
To ensure that the simulation reproduce the data well, 
a systematic comparison is made for
the missing mass distribution obtained from the 
$\Dspm$ signal region, the $\Dspm$ sideband region,
and the wrong-sign $\Dspm-\pi$ combinations both in the $\Dspm$ signal
and the $\Dspm$ sideband regions.
The ratio (Data-Monte Carlo)/Monte Carlo for all these cases 
are determined as a function of the missing mass.
We find good agreement within the errors in all cases. 
Table~\ref{tab:mc_ratio}
summarizes this result by showing the ratio integrated over the missing mass
region 1.78 to 1.87\gevcc\ for all distributions except that 
with the signal, for which the range 1.78 to 1.85\gevcc\
is used.

\begin{table}[htb]
\caption{The different data samples which can be used to determine
the background in the D$^0$ signal region.}\vspace{0.3cm}
\begin{center}
\begin{tabular}[angle=90]{|l|ccc|} \hline
Background				& Flip $\Dspms$ & $\Dspms$ Side-bands	& Side-bands flip $\Dspms$ \\ \hline\hline
1. Fake $\Dspms$ + random $\pi$		&       {\bf x}	& {\bf x}	&	{\bf x}			\\
2. Fake $\Dspms$ + correlated $\pi$     &		& {\bf x}	&				\\
3. True $\Dspms$ + random $\pi$		&	{\bf x}	&		&				\\
4. True $\Dspms$ + correlated $\pi$	&		&		&				\\ \hline
\end{tabular}
\label{tab:bgr}
\end{center}
\end{table}

\begin{table}[hbt]
\caption{The comparison of the different data samples with Monte Carlo.}
\begin{center}
\begin{tabular}[angle=90]{|l|cc|cc|} \hline 
			& \multicolumn{2}{c|}{\BDstarDs}			& \multicolumn{2}{c|}{\BDstarDsstar}		\\ \cline{2-5}
Sample type		& 
$(N_{\rm data} - N_{\rm MC})/N_{\rm MC}$
&
	$\chi^2$/{\it dof}	& 
$(N_{\rm data} - N_{\rm MC})/N_{\rm MC}$
& 
        $\chi^2$/{\it dof}	\\ \hline \hline   
$\Dspm$ Signal		& 0.051$\pm$0.025	&	1.008		&  0.103$\pm$0.057	&  1.058		\\
Flip $\Dspm$ 		&-0.043$\pm$0.031	&	0.841		& -0.041$\pm$0.064	&  0.832		\\
$\Dspm$ Sideband	& 0.006$\pm$0.018 	&	1.391		& -0.031$\pm$0.053	&  1.194		\\
Flip $\Dspm$ Sideband	& 0.015$\pm$0.021	&	1.627		&  0.084$\pm$0.069	&  1.690	 	\\
Wrong Sign 		&-0.031$\pm$0.029	&	0.987		&  0.010$\pm$0.063	&  1.088		\\
WS, $\Dspm$ Sideband 	& 0.030$\pm$0.020	&	1.311		&  0.034$\pm$0.065	&  1.487		\\ \hline
\end{tabular}
\label{tab:mc_ratio}
\end{center}
\end{table}

\section{Conclusion}

The production of $\Dspms$ at the \ups energy (and 40\mev\ below)
has been studied with the \babar\ detector. Preliminary 
measurements of branching fractions for t`inclusive production and for the 
exclusive decays \BDstarDsstar\ have been performed.
The following cross sections have been found for
production in the continuum:
\[
\sigma (e^+e^-\rightarrow \Dspm X)\cdot{\mathcal{B}}(\Dsphipi ) = 
8.29 \pm 0.41 \pm 0.69 \ {\rm pb} \;,
\]
\[
\sigma (e^+e^-\rightarrow \Dspmstar X)\cdot{\mathcal{B}}(\Dsphipi ) = 
3.48 \pm 0.39 \pm 0.38 \ {\rm pb} \;.
\]
Using the on-resonance data, the inclusive branching fraction for the B 
meson decays
\[
{\mathcal{B}}(B\rightarrow \Dspm X) = \Biggl[(11.90\pm0.30\pm1.07)\times 
\frac{3.6\pm0.9\%}{{\mathcal{B}}(\Dsphipi )}\Biggr]\% 
\]
\[
{\mathcal{B}}(B\rightarrow \Dspmstar X) = \Biggl[(6.8\pm0.7\pm0.8)\times 
\frac{3.6\pm0.9\%}{{\mathcal{B}}(\Dsphipi )}\Biggr]\%
\]
have been measured.
Finally the decays \BDstarDs\ and \BDstarDss\ have been observed using a partial
reconstruction technique and the following branching fractions have been
determined:

\begin{eqnarray*}
{\mathcal{B}} ({\rm B}\rightarrow {\rm D}_s^{+}{\rm D}^{*-}) & = &
(7.1 \pm 2.4 \pm 2.5 \pm 1.8) \times 10^{-3} \;, \\*[0.2 cm]
{\mathcal{B}} ({\rm B}\rightarrow {\rm D}_s^{*+}{\rm D}^{*-}) & = & 
(2.54 \pm 0.38 \pm 0.53 \pm 0.64)\% \;.
\end{eqnarray*}

\noindent
The results obtained are in a good agreement with previous measurements by other 
experiments. The measurement of inclusive branching fraction
of  $\Dspmstar$ from B decay has been obtained for the first time.


\section{Acknowledgements}

\input pubboard/acknowledgements.tex

\end{document}

%% file: pubboard/authors.tex
\begin{center}
\small

The \babar\ Collaboration
\bigskip

B.~Aubert,
A.~Boucham,
D.~Boutigny,
I.~De Bonis,
J.~Favier,
J.-M.~Gaillard,
F.~Galeazzi,
A.~Jeremie,
Y.~Karyotakis,
J.~P.~Lees,
P.~Robbe,
V.~Tisserand,
K.~Zachariadou
\inst{Lab de Phys.\ des Particules, F-74941 Annecy-le-Vieux, CEDEX, France}
% ...

\end{center}\newpage

%% file: pubboard/acknowledgements.tex
We are grateful for the contributions of our \pep2\ colleagues in
achieving the excellent luminosity and machine conditions
that have made this work possible.
We acknowledge support from the
Natural Sciences and Engineering Research Council (Canada),
Institute of High Energy Physics (China),
Commissariat \`a l'Energie Atomique and
Institut National de Physique Nucl\'eaire et de Physique des Particules
(France),
Bundesministerium f\"ur Bildung und Forschung
(Germany),
Istituto Nazionale di Fisica Nucleare (Italy),
The Research Council of Norway,
Ministry of Science and Technology of the Russian Federation,
Particle Physics and Astronomy Research Council (United Kingdom), the
Department of Energy (US),
and the National Science Foundation (US). In addition, individual support 
has been received from the Swiss 
National Foundation, the A. P. Sloan Foundation, the Research Corporation,
and the Alexander von Humboldt Foundation.
The visiting groups wish to thank 
SLAC for the support and kind hospitality
extended to them.